\documentclass[apj]{emulateapj}
\usepackage{amsmath}
\usepackage{apjfonts}
\usepackage{multirow}
\usepackage{array}

\newcommand{\DDir}{\relax{D\kern-.7em{/}}}

\newcommand{\be}{\begin{equation}}
\newcommand{\ee}{\end{equation}}
\newcommand{\bea}{\begin{equation*}}
\newcommand{\eea}{\end{equation*}}

\newcommand{\pr}{\partial}

\newcommand{\nin}{\relax{\in\kern-.8em{/}}}

\newcommand{\vt}{\textrm{v}}

\newcommand{\cm}{\mbox{ cm}}

\newcommand{\erg}{\mbox{ erg}}

\newcommand{\Mc}{M_{\rm c}}
\newcommand{\Menv}{M_{\rm env}}
\newcommand{\Rc}{R_{\rm c}}
\newcommand{\Teff}{T_{\rm eff}}
\newcommand{\Tph}{T_{\rm ph, RW}}
\newcommand{\LRW}{L_{\rm RW}}
\newcommand{\Tcol}{T_{\rm col}}
\newcommand{\vtf}{\vt_{\rm f}}
\newcommand{\rph}{r_{\rm ph}}

\begin{document}
\title{UV/Optical emission from the expanding envelopes of type II supernovae}
\author{Nir Sapir\altaffilmark{1,2} and Eli Waxman\altaffilmark{1}}

\altaffiltext{1}{Dept. of Particle Phys. \& Astrophys., Weizmann Institute of Science, Rehovot 76100, Israel}
\altaffiltext{2}{Plasma Physics Department, Soreq Nuclear Research Center, Yavne 81800, Israel}

\begin{abstract}
The early part of a supernova (SN) light-curve is dominated by radiation escaping from the expanding shock-heated progenitor envelope. For polytropic Hydrogen envelopes, the properties of the emitted radiation are described by simple analytic expressions and are nearly independent of the polytropic index, $n$.
This analytic description holds at early time, $t<$~few days, during which radiation escapes from shells initially lying near the stellar surface. We use numerical solutions to address two issues. First, we show that the analytic description holds at early time also for non-polytropic density profiles. Second, we extend the solutions to later times, when the emission emerges from deep within the envelope and depends on the progenitor's density profile. Examining the late time behavior of polytropic envelopes with a wide range of core to envelope mass and radius ratios, $0.1\le\Mc/\Menv\le10$ and $10^{-3}\le\Rc/R\le10^{-1}$, we find that the effective temperature is well described by the analytic solution also at late time, while the luminosity $L$ is suppressed by a factor, which may be approximated to better than $20[30]\%$ accuracy up to $t=t_{\rm tr}/a$ by $A\exp[-(at/t_{\rm tr})^\alpha]$ with $t_{\rm tr} = 13 (\Menv/M_\odot)^{3/4}(M/\Menv)^{1/4}(E/10^{51}{\rm erg})^{-1/4}$~d, $M=\Mc+\Menv$, $A=0.9[0.8]$, $a=1.7[4.6]$ and $\alpha=0.8[0.7]$ for $n=3/2[3]$. This description holds as long as the opacity is approximately that of a fully ionized gas, i.e. for $T>0.7$~eV, $t<14(R/10^{13.5}{\rm cm})^{0.55}$~d. The suppression of $L$ at $t_{\rm tr}/a$ obtained for standard polytropic envelopes may account for the first optical peak of double-peaked SN light curves, with first peak at a few days for $\Menv<1M_\odot$.
\end{abstract}

\section{Introduction}
\label{sec:intro}

During a supernova (SN) explosion, a strong radiation mediated shock wave propagates through and ejects the stellar envelope.
As the shock expands outwards, the optical depth of the material lying ahead of it decreases. When the optical depth drops below $\approx c/\vt_{\rm sh}$, where $\vt_{\rm sh}$ is the shock velocity, radiation escapes ahead of the shock and the shock dissolves. In the absence of an optically thick circum-stellar material, this breakout takes place once the shock reaches the edge of the star, producing an X-ray/UV flash on a time scale of $R/c$ (seconds to a fraction of an hour), where $R$ is the stellar radius. The relatively short breakout is followed by UV/optical emission from the expanding cooling envelope on a day time-scale. As the envelope expands its optical depth decreases, and radiation escapes from deeper shells. The properties of the breakout and post-breakout cooling emission carry unique information on the structure of the progenitor star (e.g. its radius and surface composition) and on its pre-explosion evolution, which cannot be directly inferred from observations at later time. The detection of SNe on a time scale of a day following the explosion, which was enabled recently by the progress of wide-field optical transient surveys, yielded important constraints on the progenitors of SNe of type Ia, Ib/c and II. For a recent comprehensive review of the subject see \citet[][]{WK16BOrev}.

At radii $r$ close to the stellar surface, $\delta\equiv (R-r)/R\ll1$, the density profile of a polytropic envelope approaches a power-law form,
\begin{equation}\label{eq:rho0}
  \rho_0=f_\rho \bar{\rho}_0\delta^n,
\end{equation}
with $n=3$ for radiative envelopes and $n=3/2$ for efficiently convective envelopes. Here, $\overline{\rho}_0\equiv M/(4\pi/3)R^3$ is the average pre-explosion ejecta density, $M$ is the ejecta mass  (excluding the mass of a possible remnant), and $f_\rho$ is a numerical factor of order unity that depends on the inner envelope structure \citep[see][and \S~\ref{sec:cond}, fig.~\ref{fig:frho}]{MM99,Calzavara04}. The propagation of the shock wave in this region is described by the Gandel'Man-Frank-Kamenetskii--Sakurai self similar solutions \citep{GandelMan56,Sakurai60},
\begin{equation}\label{eq:vs}
  \vt_{\rm sh}=\vt_{\rm s*} \delta^{-\beta n},
\end{equation}
with $\beta=0.191[0.186]$ for $n=3/2[3]$. The value of $\vt_{\rm s*}$ depends not only on $E$ and $M$, the ejecta energy and mass, but also on the inner envelope structure, and is not determined by the self-similar solutions alone. Based on numerical calculations, \citet{MM99} have suggested the approximation
\begin{equation}\label{eq:vstar}
  \vt_{\rm s*} \approx 1.05f_\rho^{-\beta}\sqrt{E/M}.
\end{equation}

\begin{figure}[h]
\epsscale{1} \plotone{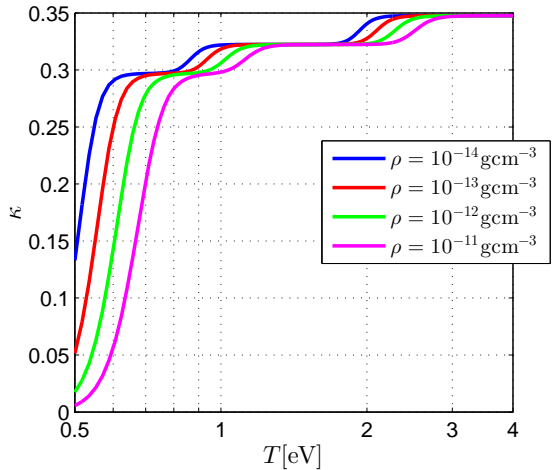}
\caption{Scattering opacity for a 30:70 (by mass) He:H mixture, at the relevant temperatures and densities. Recombination leads to opacity reduction. Similar results are obtained for solar metallicity.} \label{fig:opacity}
\end{figure}

For large Hydrogen-dominated envelopes the plasma is nearly fully ionized at early time and the opacity $\kappa$ is nearly time and space independent. In this case, the post-breakout photospheric temperature and bolometric luminosity are given, after significant envelope expansion, by \citep[][hereafter RW11]{WaxmanCampana07,RW11}
\begin{eqnarray}
\label{eq:RW11}
\Tph &=& 1.61[1.69]\left(\frac{\vt^2_{\rm s*,8.5}t_{\rm d}^2}{f_\rho M_0\kappa_{0.34}}\right)^{\epsilon_1} \frac{R_{13}^{1/4}}{\kappa^{1/4}_{0.34}}t_{\rm d}^{-1/2}\,{\rm eV}, \nonumber\\
\LRW &=&  2.0~[2.1]\times10^{42} \left(\frac{\vt_{\rm s*,8.5}t_{\rm d}^2}{f_\rho M_0\kappa_{0.34}}\right)^{-\epsilon_2} \frac{\vt^2_{\rm s*,8.5}R_{13}}{\kappa_{0.34}}\,{\rm \frac{erg}{s}},
\end{eqnarray}
where $\kappa=0.34\kappa_{0.34}{\rm cm^2/g}$, $\vt_{\rm s*}=10^{8.5}\vt_{\rm s*,8.5}{\rm cm/s}$, $M=1M_0M_\odot$, $R=10^{13}R_{13}{\rm cm}$, $\epsilon_1=0.027[0.016]$,  and $\epsilon_2=0.086[0.175]$ for $n=3/2[3]$.
This analytic description holds at times
\begin{eqnarray}\label{eq:t_limits}
   t&>&0.2\frac{R_{13}}{\vt_{\rm s*,8.5}}\max\left[0.5,\frac{R_{13}^{0.4}}{(f_\rho\kappa_{0.34}M_0)^{0.2}\vt_{\rm s*,8.5}^{0.7}} \right]\,{\rm d},  \nonumber\\
   t&<& t_\delta=3f_\rho^{-0.1}\frac{\sqrt{\kappa_{0.34}M_0}}{\vt_{\rm s*,8.5}}\,{\rm d}.
\end{eqnarray}
The first part of the lower limit, $t>R/5\vt_*$, is set by the requirement for significant expansion \citep[the shock accelerates to $>5\vt_{\rm s*}$ near the surface,][]{WK16BOrev}, while the second part is set by the requirement that the photosphere penetrates beyond the thickness of the shell at which the initial breakout takes place (where the hydrodynamic profiles deviate from the self-similar ones due to the escape of photons; see eq.~(16) of RW11). The upper limit is set by the requirement for emission from shells carrying a fraction $\delta M/M<10^{-2.5}$ of the ejecta mass, corresponding approximately to $\delta\lesssim 0.1$ (RW11).
The approximation of constant opacity holds for $T>0.7$~eV (at lower temperatures the effect of recombination becomes significant, see RW11 and fig.~\ref{fig:opacity}). At $T>0.7$~eV, the ratio of color to photospheric temperature may be approximated by (RW11) $\Tcol/T_{\rm ph}\approx 1.2$.

In RW11, $L$ and $T$ are given as functions of $E/M$ using the approximation of eq.~(\ref{eq:vstar}). Here we give $L$ and $T$ as functions of $\vt_{\rm s*}$, since this is the quantity that determines directly the emission properties, and hence constrained directly by observations, and since our numerical solutions allow us to determine $\vt_{\rm s*}$ directly, and hence to quantify the accuracy of the approximation of eq.~(\ref{eq:vstar}). Also, since our discussion is limited to the regime of time and space independent opacity, we use for $L$ the exact self-similar solution, which is available for this case \citep[][eqs. 19-20 of RW11]{Chevalier92,ChevalierFransson08}, instead of the approximate expressions (eqs. 14-15 of RW11), which differ slightly from the expressions given for $L$ in eqs.~(\ref{eq:RW11}) (in the approximate expressions, the numerical coefficients are $1.8[2.4]\times10^{42}$ and $\epsilon_2=0.078[0.15]$ for $n=3/2[3]$, and the dependence on $\vt_{\rm s*}$ is $L\propto\vt_{\rm s*}^{2-2\epsilon_2}$ instead of $\vt_{\rm s*}^{2-\epsilon_2}$, see \S~\ref{sec:LateForm}).

A comment is in place here regarding the use of eqs.~(\ref{eq:RW11}), following \cite{WaxmanCampana07} and RW11, versus the use of the rather similar results of \citet[][NS10]{NS10}. NS10 derived approximate expressions for the luminosity and temperature of both the breakout and post-breakout cooling emission. For the planar breakout phase, their estimates of $L$ and $T$ exceed those of the exact solutions \citep{SKW11Planar,KSW12Bol,SKW13Spec} by factors of a few \citep[leading to an overestimate of the optical/UV flux, which is in the Rayleight-Jeans regime at this time, by 1-2 orders of magnitude, e.g.][]{Ganot16}. For the spherical post-breakout cooling phase, the NS10 estimate of $L(t)$ is similar to that of RW11 (similar functional dependence on parameters with normalization lower[higher] by 10[40]\% compared to the exact self-similar solution of eq.~\ref{eq:RW11} for $n=3/2[3]$). The temporal and parameter dependence of the color temperature estimate of NS10 differs from that of RW11, mainly due to neglecting the bound-free absorption contribution to the opacity, which is the dominant contribution at the relevant temperatures (even for low metallicity). Due to the bound-free contribution, which is not described by Kramers' opacity law (and therefore does not follow the parameter dependence of the free-free opacity), $T_{\rm col}$ is closer to $T_{\rm ph}$ than predicted by NS10 \citep[see also the results of][showing a constant ratio of $\Tcol/T_{\rm ph}$ at late times]{1992ApJ...393..742E}. For example, for a red supergiant explosion with typical parameters ($M=15 M_\odot$, $R=500 R_{\odot}$, $E=10^{51} \erg$), the NS10 color temperature exceeds that of RW11 by 50\% at 1~d (3~eV instead of 2~eV), which implies that inferring $R$ from the observed $T_{\rm col}$ using the NS10 model would lead to an over-estimate of the radius by a factor of $\approx4$. Thus, while the approximate NS10 results may be used for an approximate description of the emission, the more accurate results of RW11 are more appropriate for inferring progenitor parameters from observations.

In this paper we use numerical solutions of the post-breakout emission to address two issues. First, we study the applicability of the analytic solution, given by eqs.~(\ref{eq:RW11}), to non-polytropic envelopes. Eqs.~(\ref{eq:RW11}) imply that $T$ is nearly independent of $n$ and essentially determined by $R$ alone, while $L$ is only weakly dependent on $n$ and determined mainly by $\vt_{\rm s*}^2R$. The near independence on $n$ suggests that the properties of the post-breakout cooling emission are nearly independent of the density profile, and therefore that eqs.~(\ref{eq:RW11}) hold also for non-polytropic envelopes. We use numerical solutions of the post-breakout emission from non-polytropic envelopes to demonstrate that this is indeed the case. In particular, we show that deviations from polytropic profiles, which are obtained by numerical stellar evolution models such as those explored by \cite{Morozova16IIP_numeric_BO}, do not lead to significant deviations from the predictions of eqs.~(\ref{eq:RW11}).

Second, we extend the analysis to $t\sim t_{\rm tr}$, when the envelope becomes transparent and emission is not limited to $\delta\ll1$ shells. At this stage, the emission is expected to depend on the envelope density structure. We present numerical solutions for progenitors composed of compact cores of radius $10^{-3}\le\Rc/R\le10^{-1}$ and mass $10^{-1}\le\Mc/M\le10^{1}$, surrounded by extended H-dominated $n=3/2$ and $n=3$ polytropic envelopes of mass $\Menv=M-\Mc$, and provide analytic approximations describing the deviation from eqs.~(\ref{eq:RW11}) at late time (in our numerical calculations the entire core mass $\Mc$ is ejected; the results are not sensitive to the presence of a remnant).

As explained in \S~\ref{sec:LateForm}, $T_{\rm ph}$ and $L$ are given at $t\gg R/\vt_{\rm s*}$ by
\begin{eqnarray}
\label{eq:T_L_dimensions}
T_{\rm ph} &=& f_T\left(\xi,c/\vt_{\rm s*},\alpha_i\right) \left(\frac{R}{\kappa t^2}\right)^{1/4}, \nonumber\\
L &=&  f_L\left(\xi,c/\vt_{\rm s*},\alpha_i\right) \left(\frac{c\vt_{\rm s*}^2R}{\kappa}\right),
\end{eqnarray}
where $f_T$ and $f_L$ are $R$-independent dimensionless functions of the dimensionless variable $\xi\equiv c\vt_{\rm s*}t^2/\kappa \Menv$, of $c/\vt_{\rm s*}$ and of a set of dimensionless parameters $\alpha_i$ determining the progenitor structure ($n,\Mc/M$,$\Rc/R$). We use our numerical calculations to determine $f_L$ and $f_T$ and to study their dependence on $\alpha_i$.

Our approach is complementary to that using numerical calculations to derive the post breakout emission properties for progenitor structures ($\alpha_i$), which are determined by stellar evolution calculations under specific assumptions regarding processes (like convection and mass loss), for which a basic principles theory does not yet exist. Uncertainties in $\alpha_i$ arise due to the absence of such a theory, as reflected in the varying results obtained by different numerical calculations. Our analysis enables us to explore a wide range of progenitor parameters, to determine which characteristics of the emission are not sensitive to uncertainties in $\alpha_i$ (due to uncertainties in stellar evolution models), and to determine the dependence on $\alpha_i$ of the characteristics which are sensitive to these uncertainties.

This paper is organized as follows. The equations solved and the initial conditions used are described in \S~\ref{sec:EqCond}. We solve the radiation hydrodynamics equations, using the diffusion approximation with constant opacity. The general form of the solutions at $t\gg R/\vt_{\rm s*}$ (eq.~\ref{eq:T_L_dimensions}) is derived in \S~\ref{sec:LateForm}. The numerical results are presented in \S~\ref{sec:results}. A summary of the analytic formulae, which provide an approximate description of the post-breakout cooling emission, is given in \S~\ref{sec:analytic}. Double-peaked SN light curves are discussed in \S~\ref{sec:2peak}. In \S~\ref{sec:discussion} our results are summarized and discussed, with a focus on the implications for what can be learned about the progenitors from post-breakout emission observations.

\section{Equations and initial conditions}
\label{sec:EqCond}

\subsection{Equations}
\label{sec:eq}

We consider a spherically symmetric non-relativistic flow of an ideal fluid, with pressure dominated by radiation and approximating radiation transport by diffusion with constant opacity. Using Lagrangian coordinates, labeling a fluid element by the mass $m$ enclosed within the radius $r$ at which it is located, the radiation-hydrodynamics equations describing the evolution of the radius $r$, the velocity $\vt$, and the energy density $e$ of a fixed fluid element are
\begin{align}
&\pr_t r=\vt, \label{eq:VelocityDef}\\
&\pr_{t}\vt=-4\pi r^2\pr_m p,\label{eq:MomentumCons}\\
&\pr_t(e/\rho)=-\pr_m (4\pi r^2 j)-p\pr_m(4\pi r^2\vt),\label{eq:EnergyCons}\\
&j=-\frac{c}{3\kappa}4\pi r^2 \pr_m e,  \label{eq:CurrDiff}\\
&p=e/3, \label{eq:EOS}
\end{align}
where $\rho=(4\pi r^2)^{-1}\pr_r m$ is the density, $j$ is the energy flux (energy current density) and $p$ is the radiation pressure. The optical depth is given by $\tau(r)=\int_{r}^{\infty} \kappa\rho dr'$. Gravity does not affect the flow of the ejecta significantly, and is therefore neglected. The equations were solved by a standard leap-frog method on a staggered-mesh, with a fully-implicit solver for the energy equation.

A stationary inner boundary condition, $\vt=0$ and $j=0$, and a free surface outer boundary condition, $\pr_{t}\vt=(\kappa/c)j$ and $e=0$, were imposed at $m=0$ and at $m=M$, respectively. The bolometric luminosity is not sensitive to the exact choice of the boundary condition at $m=M$, since it is determined by the diffusion through the optically thick layers \citep[see][]{SKW11Planar}. Invoking an ad-hoc radiation flux-limiter is not required in our numeric calculations, since at $\tau \ge 1$ the flux is naturally limited at all times to $j/ce<0.5$ (at $\tau < 1$ the flux is determined by the flux at $\tau \approx 1$, approximately satisfying $\nabla j=0$, and the energy density is unimportant). This also justifies our choice of the outer boundary conditions \citep[see also][]{1992ApJ...393..742E}.

\subsection{Initial conditions}
\label{sec:cond}

\begin{figure}[h]
\epsscale{1} \plotone{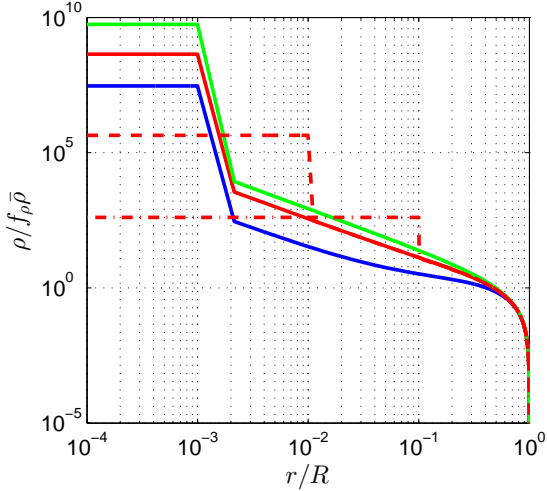}
\caption{Initial density profiles as a function of radius for polytropic $n=3/2$ envelopes with various $\Mc/\Menv$ and $\Rc/R$ values. Blue, red and green lines correspond to $\Mc/\Menv=0.1,1,10$, respectively. Solid, dashed and dash-dotted lines correspond to $\Rc/R=10^{-3},10^{-2},10^{-1}$, respectively.} \label{fig:initialdensity}
\end{figure}

\begin{figure}[h]
\epsscale{1} \plotone{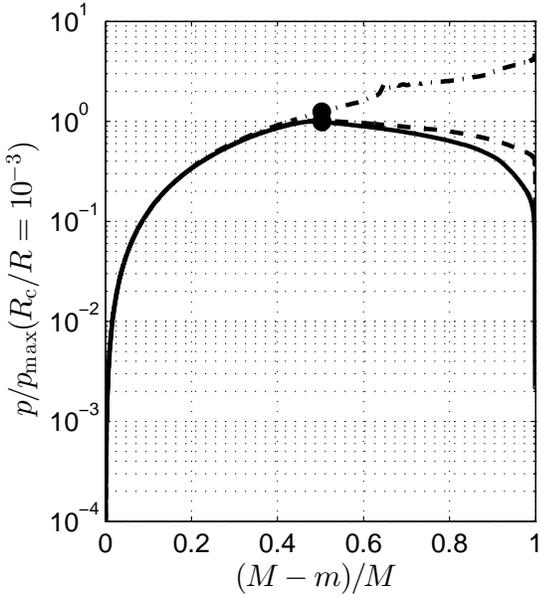}
\caption{The ratio of $p(m)$ obtained for $\Rc/R=10^{-3},10^{-2},10^{-1}$ (solid, dashed, dashed-dot) to the maximum pressure obtained for $\Rc/R=10^{-3}$ at $t= R/\vt_{\rm s*}$, for $\Mc/\Menv=1$ and $n=3/2$. Circles denote the core's location.} \label{fig:p_profile}
\end{figure}

\begin{figure}[h]
\epsscale{1} \plotone{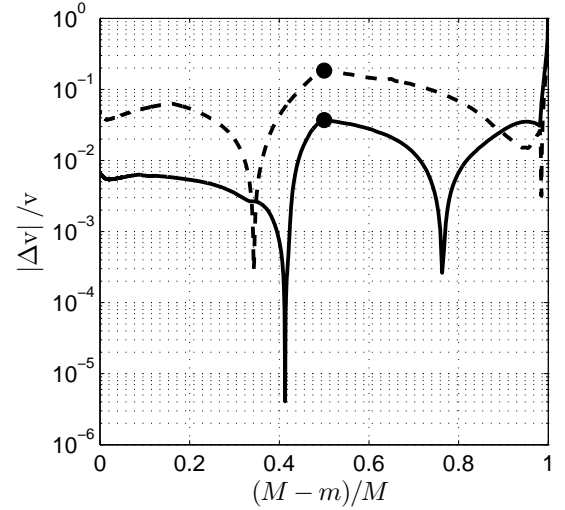}
\caption{The absolute value of the fractional difference between $\vt(m)$ obtained for $\Rc/R=10^{-2},10^{-1}$ (solid, dashed) and $\vt(m)$ obtained for $\Rc/R=10^{-3}$ at $t= R/\vt_{\rm s*}$, for $\Mc/\Menv=1$ and $n=3/2$. Circles denote the core's location.} \label{fig:v_profile}
\end{figure}

In order to study the late time, $t>t_\delta$, behavior, we consider progenitors of radius $R$ and mass $M$, composed of a uniform density core of mass $\Mc$ and radius $\Rc\ll R$, surrounded by a polytropic envelope in hydrostatic equilibrium. At $t=0$ an energy $E$ is uniformly distributed within $r<\Rc/3$ to initiate the "explosion". In these calculations the entire mass $M$ is ejected, hence $M$ represents the ejecta mass (i.e. excluding the mass of a remnant). We consider $n=3/2$ and $n=3$ envelopes, a wide range of core to envelope masses, $0.1<\Mc/(M-\Mc)<10$, a wide range of core to envelope radii, $10^{-3}<\Rc/R<10^{-1}$, and a wide range of radii, $10^{12}\cm<R<5\times 10^{13} \cm$.
Figure \ref{fig:initialdensity} shows the initial density profiles for several $\Mc/(M-\Mc)$ and $\Rc/R$ values, while figures~\ref{fig:p_profile} and~\ref{fig:v_profile} show the pressure and velocity profiles obtained at $t\approx R/\vt_{\rm s*}$.  At late times, $t>5R/\vt_{\rm s*}$, the pressure and velocity profiles are not sensitive to the value of $\Rc/R$ for $\Rc/R<0.1$ (the fractional variations between the $\Rc/R=0.1,0.01$ and $\Rc/R=10^{-3}$ solutions are $\lesssim10,1\%$; see also figs.~\ref{fig:Lq32} and~\ref{fig:Lq3}). In what follows we present results for $\Rc/R=10^{-3}$, unless specifically stated otherwise.

\begin{figure}[h]
\epsscale{1} \plotone{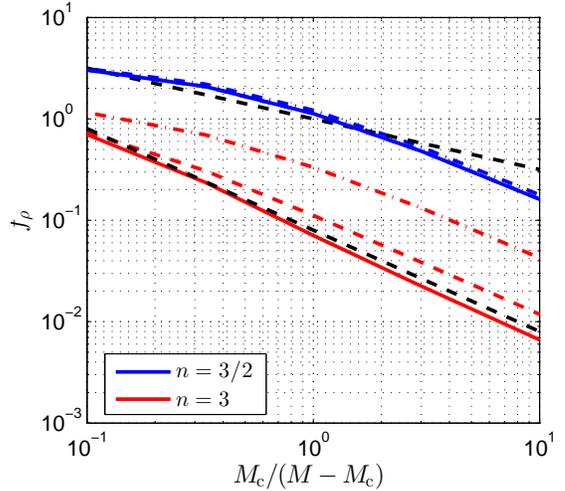}
\caption{$f_\rho$, derived from the numerical profiles using eq.~(\ref{eq:rho0}), as a function of $\Mc/(M-\Mc)=\Mc/M_{\rm env}$ for $n=3/2,3$. Solid, dashed, and dash-dotted lines correspond to $\Rc/R=10^{-3},10^{-2}$ and $10^{-1}$, respectively. Black dashed lines show the approximations $f_\rho=(\Menv/\Mc)^{1/2}$ and $f_\rho=0.08(\Menv/\Mc)$ for $n=3/2$ and $n=3$.\label{fig:frho}}
\end{figure}

We note, that the convergence of the initial density profiles to the $\Rc/R=10^{-3}$ profile is slower for $n=3$ compared to $n=3/2$ envelopes, see fig.~\ref{fig:frho}. This, combined with the fact that the core radii of blue supergiants with radiative, $n=3$, envelopes may reach $\Rc/R\sim$~a few \%, implies that the value of $f_\rho$ appropriate for such progenitors may depend on $\Rc/R$ and not only on $\Mc/M$. Although the dependence of the properties of the emission on $f_\rho$ is weak, the sensitivity of $f_\rho$ to $\Rc/R$ for large $\Rc/R$ and $n=3$ should be kept in mind (e.g. when inferring $E/M$ from $\vt_*$, see eq.~\ref{eq:vstar} and \S~\ref{sec:discussion}).

\begin{figure}[h]
\epsscale{1} \plotone{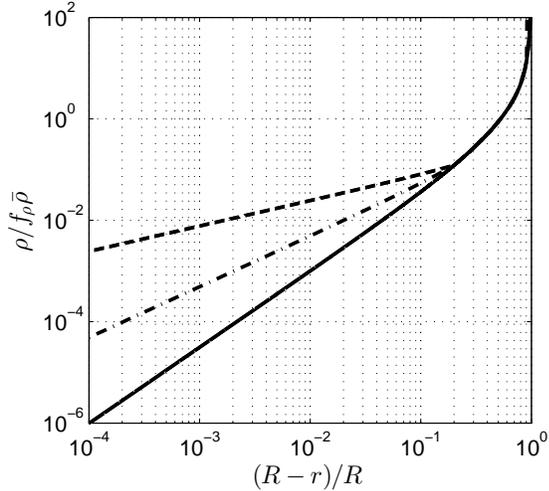}
\caption{Polytropic (solid) and modified (dashed- $\tilde{n}=0.5$, dash-dotted- $\tilde{n}=1$) density profiles used in the calculations. } \label{fig:ModDen}
\end{figure}

In order to study the dependence of the early time, $t<t_\delta$, behavior on deviations from polytropic profiles, we solve the radiation hydrodynamics equations for modified initial density profiles, where the density profile of an $n=3/2$ envelope is modified at the outer radii, $r>0.8R$, to $\rho_0\propto \delta^{\tilde{n}}$ with $\tilde{n}=0.5,1$ (keeping a continuous density at $r=0.8R$, see fig.~\ref{fig:ModDen}). These modified profiles span the range of density profiles obtained by \citet{Morozova16IIP_numeric_BO} using the MESA and KEPLER stellar evolution codes at the relevant radii range (see their figure 2). We note, that the KEPLER profiles are not described at the outermost $\delta M/M< 10^{-3}$ shells by a smooth power-law of the form given above. While this deviation may affect the breakout emission, it does not affect the post breakout emission discussed here, produced by deeper shells (see eq.~\ref{eq:t_limits} above and eq.~6 of RW11).

\section{The general form of the solutions at $t\gg R/\vt_{\rm s*}$}
\label{sec:LateForm}

The functional dependence of the solutions on $R$ at $t\gg R/\vt_{\rm s*}$ may be inferred as follows. Let us compare the solution obtained for some initial conditions, $\rho_0(r)$, $p_0(r)$, and $\vt_0(r)=0$, to a solution obtained for modified initial conditions, $\tilde{\rho}_0(r)=X^{-3}\rho_0(r/X)$, $\tilde{p}_0(r)=X^{-3}p_0(r/X)$, $\tilde{\vt}_0(r)=0$. $E$, $M$ and the initial progenitor structure ($n,\Mc/M,\Rc/R$) are the same for both solutions, while $R$ is larger by a factor $X$ for the modified initial conditions.

Let us consider first the evolution neglecting photon diffusion. Each fluid element is accelerated first as it is shocked by the shock wave, and then as the fluid expands and converts its internal energy to kinetic energy. The latter stage of acceleration ends at $t\sim R/\vt_{\rm s*}$, and the fluid reaches an asymptotic velocity profile, $\vt(m,t)=\vtf(m)$, at $t\gg R/\vt_{\rm s*}$. It is straight forward to verify that, neglecting diffusion, the shock velocity profiles of both solutions are the same, $\tilde{\vt}_{\rm sh}(m)=\vt_{\rm sh}(m)$, and the asymptotic velocity profiles of both solutions are the same, $\tilde{\vt}_{\rm f}(m)=\vtf(m)$. $\tilde{\vt}_{\rm f}(m)=\vtf(m)$ further implies that the density profiles at $t\gg R/\vt_{\rm s*}$ are also the same, $\tilde{\rho}(m,t)=\rho(m,t)$.

Consider next the pressure. Neglecting diffusion, conservation of entropy implies that the pressure of a fluid element $m$, $p(m,t)$ is related to the pressure it reached at shock passage, $p_{\rm sh}(m)=(6/7)\rho_0(m)\vt^2_{\rm sh}(m)$, by
\begin{equation}\label{eq:pmt}
  p(m,t)=\left[\frac{\rho(m,t)}{7\rho_0(m)}\right]^{4/3}p_{\rm sh}(m)\propto \left(\frac{\rho}{\rho_0}\right)^{1/3}\rho\vt_{\rm sh}^2
\end{equation}
(note that the shock compresses the fluid density by a factor of 7). Noting that $\tilde{\rho}_0(r)=X^{-3}\rho_0(r/X)$, $\tilde{\vt}_{\rm sh}(m)=\vt_{\rm sh}(m)$, and $\tilde{\rho}(m,t)=\rho(m,t)$ we find that $\tilde{p}(m,t)=Xp(m,t)$.

Thus, increasing $R$ by a factor $X$, keeping $E$ and $M$ fixed, does not change the asymptotic, $t\gg R/\vt_{\rm s*}$, velocity and density profiles and increases the pressure everywhere by a factor $X$. Photon diffusion leads to modifications of the density and velocity profiles only at the outermost shells, from which radiation may escape at $t<R/\vt_{\rm s*}$. This does not affect the solution for the escaping radiation at late time.

Since the asymptotic pressure and energy density are proportional to $R$, we must have $T\propto R^{1/4}$ and $L\propto R$. This implies that $T_{\rm ph}$ and $L$ are given at $t\gg R/\vt_{\rm s*}$ by $T_{\rm ph}=f_T (R/\kappa t^2)^{1/4}$ and $L=f_L (c\vt_{\rm s*}^2R/\kappa$), where $f_T$ and $f_L$ are $R$-independent dimensionless functions, which can depend only on dimensionless variables constructed of $t$ and $\kappa/c$ (which appears in the equations), and of the parameters determining the initial and boundary conditions (of which three are dimensional, $M$, $\vt_{\rm s*}$, $c$). We may choose the dimensionless parameters as $\xi\equiv \vt_{\rm s*}t^2/(\kappa/c) \Menv$, $c/\vt_{\rm s*}$ and a set of dimensionless parameters determining the progenitor structure, $\{\alpha_i\}=\{n,\Mc/M,\Rc/R\}$.

The dimensional parameter $c$ affects the solution of the diffusion equation through the boundary condition for the escaping flux set at $\tau\sim1$, beyond which the diffusion approximation does not hold. We expect the dependence on the choice of boundary condition to be weak, see \S~\ref{sec:eq}, and hence $f_L$ to depend on $\xi$ and $\{\alpha_i\}$ only. On the other hand, the location of the photosphere depends on $\kappa$, rather than on $\kappa/c$, and is therefore given by $\rph=f_{\rm ph}(\tilde{\xi},\alpha_i)\vt_{\rm s*}t$, where $\tilde{\xi}=\xi \vt_{\rm s*}/c= \vt^2_{\rm s*}t^2/\kappa \Menv$.

\section{Results}
\label{sec:results}

We discuss in \S~\ref{sec:hydro} some aspects of the hydrodynamic behavior of the solutions, comparing our results to those of earlier work. Our main results, regarding the properties of the emitted radiation, are presented in \S~\ref{sec:rad}.

At the early stages of the explosion, when the shock propagates through the inner, $\tau\gg1$, parts of the envelope, radiation diffusion does not affect the flow significantly. This enables one to reduce the calculation time by dividing the calculations into two stages. At the first stage, a pure hydrodynamic calculation of the explosion is carried out, neglecting radiation diffusion. At the second stage, a full radiation-hydrodynamics calculation is carried out, with initial conditions given by the hydrodynamic profiles obtained following significant expansion of the ejecta.

For the first stage we used a grid which is uniform in radius within the stellar core, and logarithmically spaced in radius within the envelope, with increased resolution towards the stellar surface. This choice has shown the fastest convergence. A nominal grid of 4200 cells was used, with 200 cells within the core and 4000 cells within the envelope, with a minimal spacing of $10^{-4}R$. Comparing the results with those of calculations with 2100 and 8400 envelope cells, keeping a similar 1/20 ratio of core to envelope cell numbers, we find that all the important envelope parameters ($\vt_{\rm s*}$, $\vt(m,t> R/\vt_{\rm s*}$, $\rho(m,t> R/\vt_{\rm s*}$ and $e(m,t> R/\vt_{\rm s*}$) are converged in the nominal calculations to within 2\%.

For the second stage of the calculation, a new grid was chosen, with logarithmic spacing in optical depth and resolution increasing towards the photosphere. A nominal grid of 1000 cells was used, with a minimal optical depth of 0.1. Comparing the results with those of calculations with 2000 and 4000 cells, we find that the bolometric luminosity and effective temperature are converged in the nominal calculations to within 2\% and 0.5\% respectively.
The total energy is conserved to within 1\% in all calculations.

\subsection{Hydrodynamics}
\label{sec:hydro}

\begin{figure}[h]
\epsscale{1} \plotone{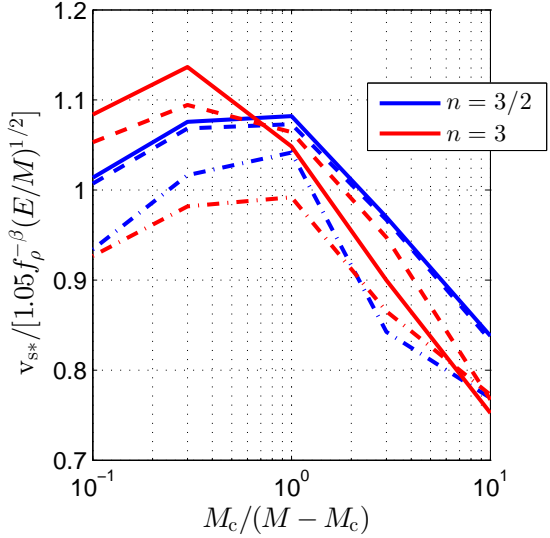}
\caption{$\vt_{\rm s*}$, derived from the numerical profiles using eq.~\eqref{eq:vs} and normalized to the approximation of eq.~\eqref{eq:vstar}, as a function of $\Mc/\Menv$ for $n=3/2,3$. Solid, dashed, and dash-dotted lines correspond to $\Rc/R=10^{-3},10^{-2}$ and $10^{-1}$, respectively.} \label{fig:vst}
\end{figure}

Figures~\ref{fig:frho} and~\ref{fig:vst} present the dependence on $\Mc/(M-\Mc)=\Mc/M_{\rm env}$ of $f_\rho$ and of $\vt_{\rm s*}$ , normalized to the approximation suggested by \citet{MM99}, eq.~(\ref{eq:vstar}).
We find that the approximation of eq.~(\ref{eq:vstar}) holds to better than 10\% for $0.3<\Mc/M_{\rm env}<3$. The dependence of $f_\rho$ on $\Mc/M_{\rm env}$, approximately given by $f_\rho=(\Menv/\Mc)^{1/2}$ and $f_\rho=0.08(\Menv/\Mc)$ for $n=3/2$ and $n=3$, implies that, as expected, the relation between $\vt_{\rm s*}$ and $E/M$, which characterizes the bulk ejecta velocity, depends on the ejecta structure. In the absence of detailed information on the structure, $E/M$ may be inferred from $\vt_{\rm s*}$, which may be determined by early UV observations through eqs.~(\ref{eq:RW11}), by $E/M=0.9[0.3]\vt_{\rm s*}^2$ for $n=3/2[3]$ with $5[30]\%$ accuracy for $0.3<\Mc/M_{\rm env}<3$. Conversely, a comparison of $\vt_{\rm s*}$, determined by early UV observations, and $E/M$, determined by other late time observations (e.g. spectroscopic ejecta velocity), will constrain the progenitor structure.

\begin{figure}[h]
\epsscale{1} \plotone{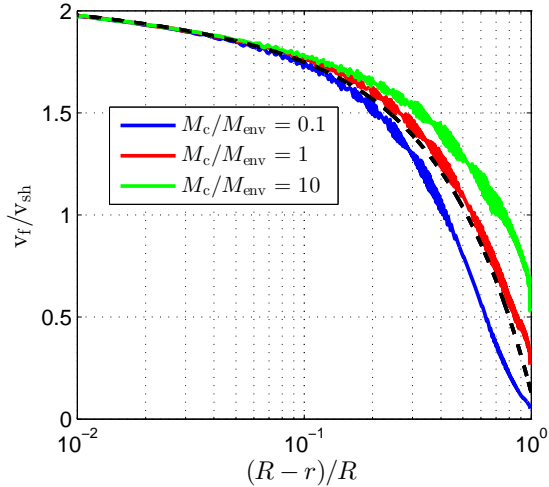}
\caption{The ratio $f_\vt=\vtf(m)/\vt_{\rm sh}$ as a function of $\delta$ for different values of $\Mc/\Menv$, for $n=3/2$. Blue, red and green curves correspond to $\Mc/\Menv=0.1,1,10$, respectively. The analytic approximation of equation~6 of \citet{MM99} is shown as a black dashed line. The "noise" in the numerical curves reflects the inaccuracy in the numerical derivative of the shock's position as a function of time. \label{fig:f_v_n15}}
\end{figure}

\begin{figure}[h]
\epsscale{1} \plotone{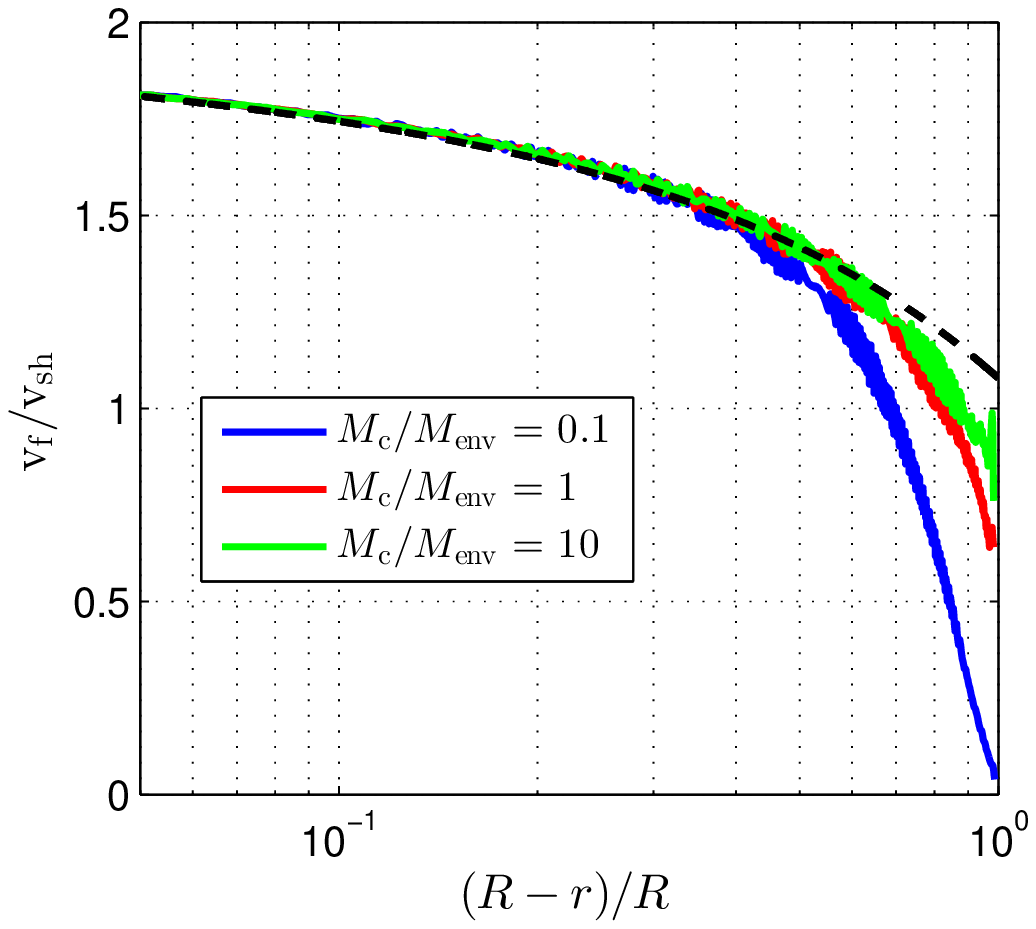}
\caption{The same as fig.~\ref{fig:f_v_n15}, for $n=3$. \label{fig:f_v_n3}}
\end{figure}

Figures~\ref{fig:f_v_n15} and \ref{fig:f_v_n3} present the ratio of the final velocity, to which each fluid element is accelerated, to the velocity of the shock passing through this fluid element, $f_\vt(m)=\vtf/\vt_{\rm sh}$, for $n=3/2$ and $n=3$ respectively. We find that the spherical correction to the planar self-similar dynamics, described by $f_\vt$ and given by eq.~26 of \citet{MM99}, is accurate for the outer parts of the ejecta. In the inner parts, where the flow deviates from the self-similar solution, $f_\vt(m)$ also deviates from that given by \citet{MM99} and depends on the detailed structure.

\subsection{Radiation}
\label{sec:rad}

\begin{figure}[h]
\epsscale{1} \plotone{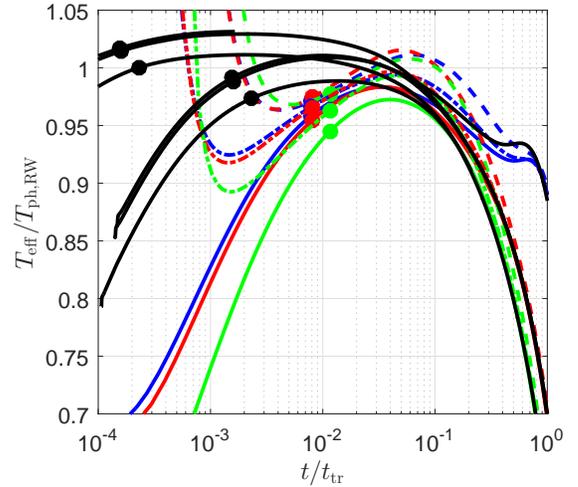}
\caption{A comparison of $\Teff$ obtained in the numerical calculations with the analytic model of eq.~(\ref{eq:RW11}), for $n=3/2$. Blue, red and green curves correspond to $R=5\times10^{13}$~cm with $\Mc/\Menv=0.1,1,10$, respectively. Solid lines correspond to polytropic envelopes, while dashed and dash-dotted lines correspond to modified envelopes with $\tilde{n}=0.5,1$ respectively. Circles denote $t=R/5\vt_{\rm s*}$, the time beyond which the solution is expected to be described by eq.~\eqref{eq:RW11}. Black curves show the results for $R=1\times10^{13}$~cm ($R/5\vt_{\rm s*}\approx2\times10^{-3}t_{\rm tr}$) and $R=1\times10^{12}$~cm ($R/5\vt_{\rm s*}\approx2\times10^{-4}t_{\rm tr}$) for $\Mc/\Menv=1$ (top curves) and $10$ (bottom curves). Note that the validity of the model is limited to times at which $T_{\rm ph}>0.7$~eV. \label{fig:Teff32}}
\end{figure}

\begin{figure}[h]
\epsscale{1} \plotone{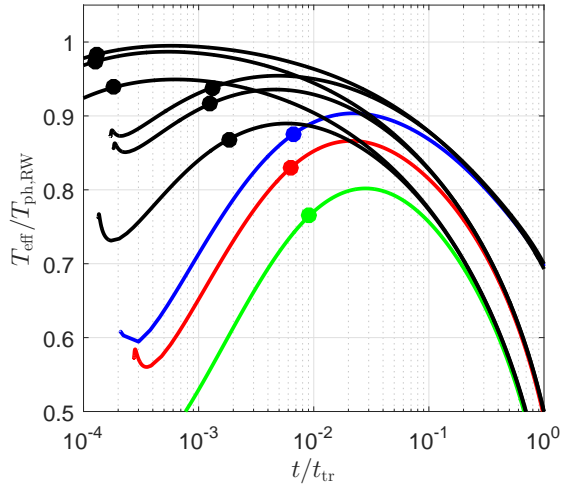}
\caption{The same as figure \ref{fig:Teff32}, for $n=3$ (not including modified density profiles).\label{fig:Teff3}}
\end{figure}

\begin{figure}[h]
\epsscale{1} \plotone{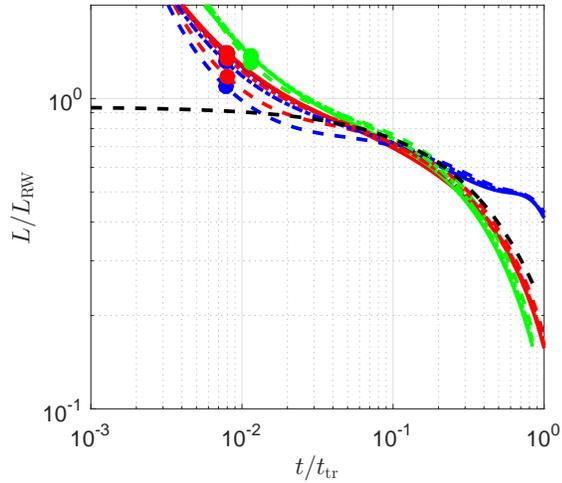}
\caption{The ratio of the luminosity obtained in the numerical calculation, $L$, to the luminosity $\LRW$ given by the analytic approximation of eq.~\eqref{eq:RW11}, for $n=3/2$. Line colors and types correspond to the same parameter choices as in figure \ref{fig:Teff32}. The thick black dashed line shows the fitting formula of eq.~(\ref{eq:Lq}). The light curves of modified envelopes with $\tilde{n}=0.5,1$ (dashed and dash-dotted lines) nearly coincide with those of unmodified $n=3/2$ envelopes. \label{fig:Lqmodified32}}
\end{figure}

\begin{figure}[h]
\epsscale{1} \plotone{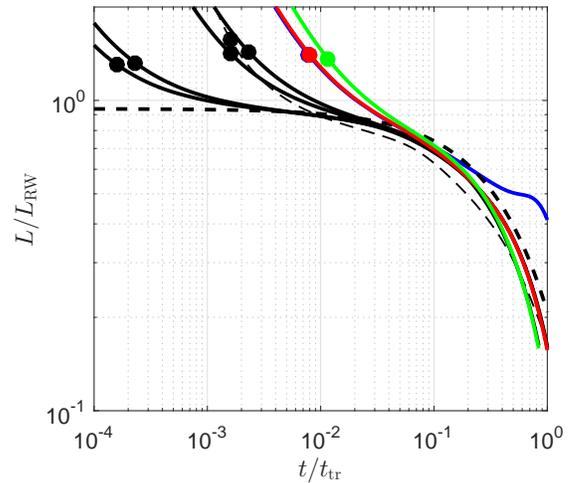}
\caption{The ratio of the luminosity obtained in the numerical calculation, $L$, to the luminosity $\LRW$ given by the analytic approximation of eq.~\eqref{eq:RW11}, for $n=3/2$. Line colors and types correspond to the same parameter choices as in figure \ref{fig:Teff32}. The thin black dashed line corresponds to a progenitor with a large core radius, $\Rc/R=0.1$, with $R=10^{13}$~cm and $\Mc/M=1$. The thick black dashed line shows the fitting formula of eq.~(\ref{eq:Lq}). \label{fig:Lq32}}
\end{figure}

\begin{figure}[h]
\epsscale{1} \plotone{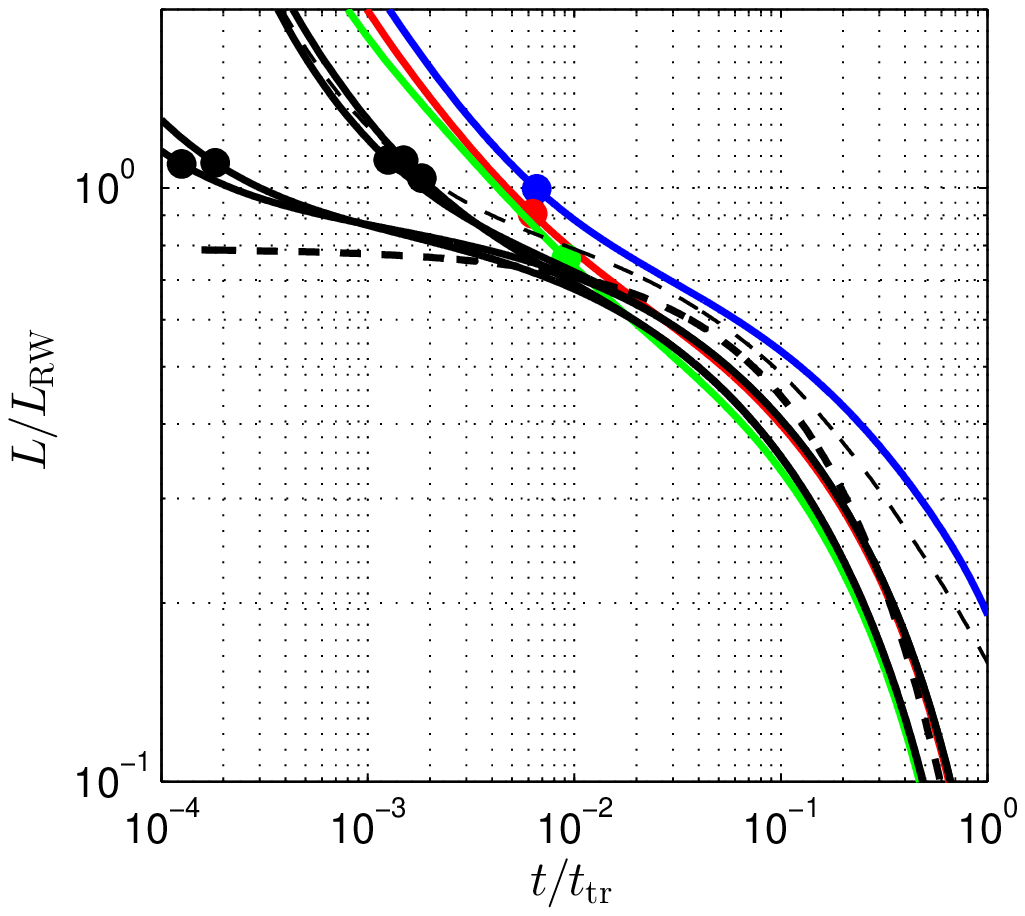}
\caption{The same as fig.~\ref{fig:Lq32}, for $n=3$. \label{fig:Lq3}}
\end{figure}

\begin{figure}[h]
\epsscale{1} \plotone{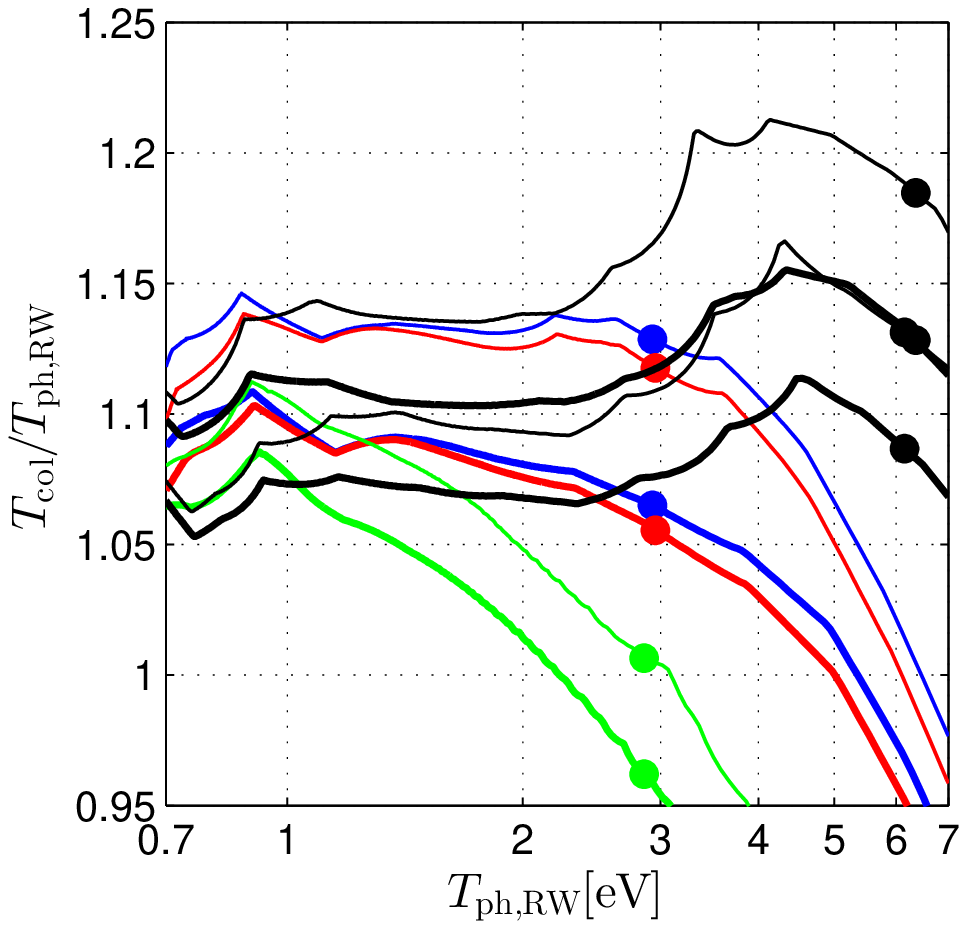}
\caption{The ratio $\Tcol/\Tph$ as a function of $\Tph$ for polytropic envelopes with $n=3/2$. $\Tph$ is given by eq.~(\ref{eq:RW11}) and $\Tcol$ is calculated from the numerical radiation pressure at the thermalization depth (see text). Blue, red and green curves correspond to $R=5\times10^{13}$~cm with $\Mc/\Menv=0.1,1,10$, respectively. Black curves show the results for $R=1\times10^{12}$~cm for $\Mc/\Menv=1$ (top curves) and $10$ (bottom curves). Circles denote $t=R/5\vt_{\rm s*}$, the time beyond which the solution is expected to be described by eq.~\eqref{eq:RW11}. Thick lines correspond to solar metallicity opacity, thin lines to 0.1 solar metallicity. The "noise" in the numerical curves reflects the finite resolution of the determination of the (non-monotonic) temperature dependence of the opacity (obtained through interpolations within the opacity tables). \label{fig:Tc32}}
\end{figure}

\begin{figure}[h]
\epsscale{1} \plotone{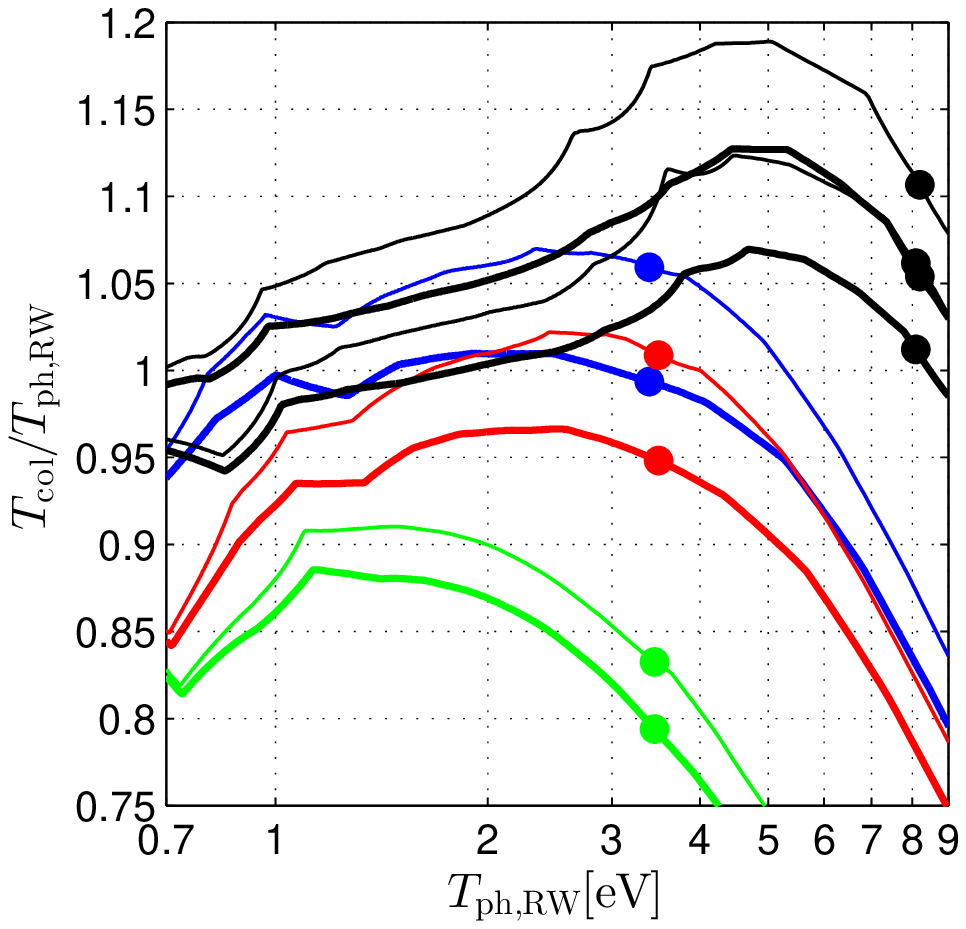}
\caption{The same as figure \ref{fig:Tc32}, for $n=3$.} \label{fig:Tc3}
\end{figure}

\begin{figure}[h]
\epsscale{1} \plotone{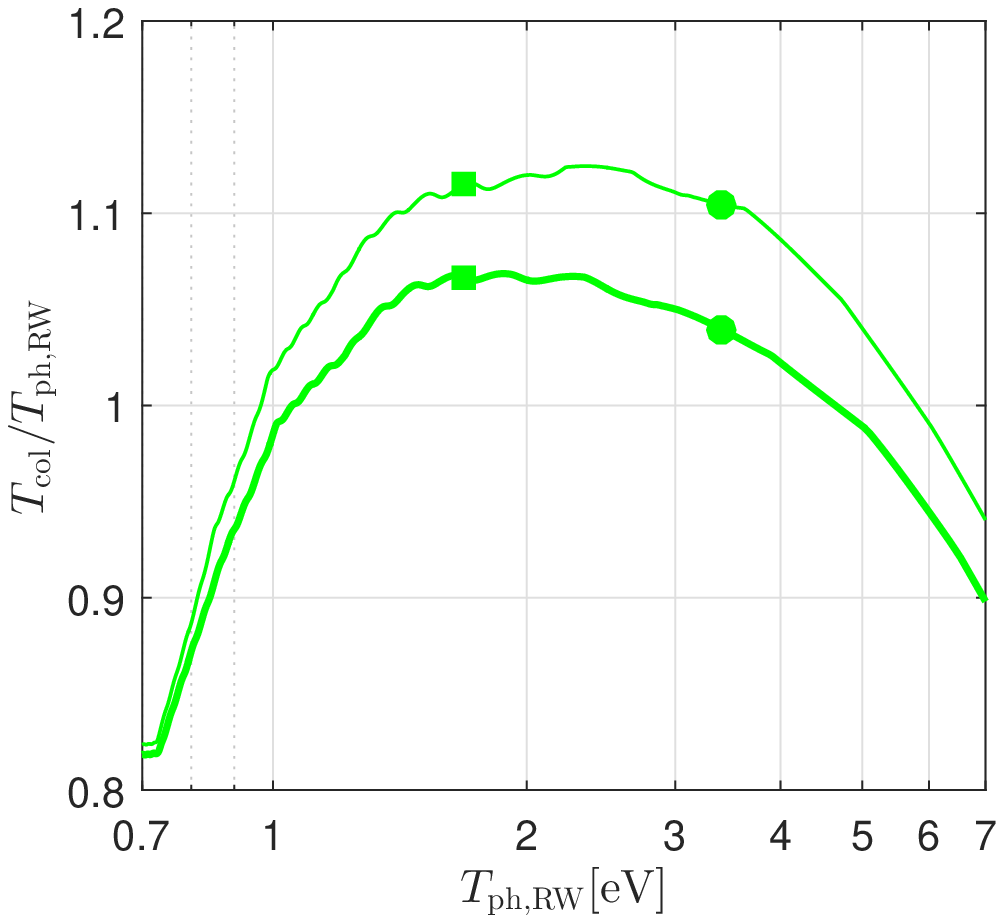}
\caption{$\Tcol/\Tph$ as a function of $\Tph$ for an explosion with $n=3/2$, $R=5\times 10^{13}~{\rm cm}$, $\Mc/\Menv=10$ and $\Menv=1M_\odot$. Thick lines correspond to solar metallicity opacity, thin lines to 0.1 solar metallicity. Circles denote $t=R/5\vt_{\rm s*}$, the time beyond which the solution is expected to be described by eq.~\eqref{eq:RW11}. Squares denote the time $t=0.1t_{\rm tr}$. \label{fig:Tc32_Menv1}}
\end{figure}

Figures \ref{fig:Teff32}-\ref{fig:Lq3} present the results of our numerical calculations for $\Teff$ and $L$, where $\Teff$ is defined through $L=4\pi \rph^2\sigma \Teff^4$ and the photospheric radius is determined by $\tau(\rph)=1$. The figures show the ratio between $\Teff$ and $L$ obtained numerically and the analytic results of eqs.~(\ref{eq:RW11}), with $f_\rho$ and $\vt_{\rm s*}$ determined from the numerical solutions, and with time normalized to
\begin{equation}\label{eq:tstar}
  t_{\rm tr}=\left(\frac{\kappa \Menv}{8\pi c\vt_{\rm s*}}\right)^{1/2}=19.5\left(\frac{\kappa_{0.34} M_{\rm env,0}}{\vt_{\rm s*,8.5}}\right)^{1/2}{\rm d},
\end{equation}
such that $t/t_{\rm tr}=\sqrt{8\pi\xi}$ (see eq.~\ref{eq:T_L_dimensions}). $t_{\rm tr}$ is the time at which the envelope is expected to become transparent, i.e. satisfying $\tau\sim \kappa \Menv/4\pi \vt^2t^2=c/\vt$, noting that $\vtf\sim2\vt$.
Circles denote the time $t=R/5\vt_{\rm s*}$, after which the approximation of significant expansion is expected to hold. Results for polytropic envelopes are presented in solid lines, and for modified density profiles in dashed (dash-dotted) lines for $\tilde{n}=0.5(1)$. The figures clearly demonstrate that, as expected, the properties of the cooling envelope emission are not sensitive to the details of the density profile near the stellar surface. It should be noted here, that the photosphere lies within the layers of modified initial density at all times shown.

While \citet{Morozova16IIP_numeric_BO} show that the prescription of identifying the "luminosity shell" with $\tau=c/\vt$ is inappropriate for shallow density profiles, our numerical results show that eqs.~\ref{eq:RW11} with $n=3/2$ describe $L$ and $T$ also for $n<3/2$ envelopes. This can be explained by the weak $n$ dependence of both the luminosity and the temperature \citep[note that the self-similar solutions are not physical for $n \lesssim 0.9$, where their luminosity diverges at $r \rightarrow 0$,][]{Chevalier92}.

At late time, $t> t_\delta$, radiation emerges from inner layers, the properties of which are not well approximated by the self-similar solution determined by eqs.~(\ref{eq:rho0}) and ~(\ref{eq:vs}) (with post-breakout acceleration given by a fixed value of $f_\vt=\vt_f/\vt_{\rm s}$, see figs.~\ref{fig:f_v_n15} and~\ref{fig:f_v_n3}). This leads to a suppression of the luminosity below the nearly time independent luminosity given by eq.~(\ref{eq:RW11}), which is valid for $t< t_\delta$. The suppression of $L$ may be approximately described by the analytic expression (obtained by fitting to the numerical results)
\begin{equation}\label{eq:Lq}
  L/\LRW= A\exp\left[-\left(\frac{at}{t_{\rm tr}}\right)^\alpha\right],
\end{equation}
with $A=0.94[0.79]$, $a=1.67[4.57]$ and $\alpha=0.8[0.73]$ for $n=3/2[3]$. This approximation holds to better than $20[30]\%$ up to $t=t_{\rm tr}/a$ for $n=3/2[3]$. The temperature is affected less strongly. For $n=3/2$, $T_{\rm eff}(t<0.5t_{\rm tr})$ is smaller than the self-similar result by $<10\%$ for $\Menv/M_c>>1$ and by $<20\%$ for $\Menv/M_c\le1$.

Figures~\ref{fig:Tc32} and~\ref{fig:Tc3} present the numerical results for the color temperature, $\Tcol$, defined as the temperature of the plasma obtained in the numerical calculations (assuming local thermal equilibrium) at the "thermalization depth" $r_{\rm ther}$, from which photons may diffuse to the photosphere without being absorbed. This radius is estimated as the radius for which the product of scattering and absorption optical depths equals unity, $\tau_{\rm sct}\tau_{\rm abs}\approx1$ \citep{Mihalas84}, approximately determined by (see RW11)
\begin{equation}\label{eq:r_therm}
3(r_{\rm ther}-\rph)^2\kappa_{\rm sct}(r_{\rm ther})\kappa_{\rm abs}(r_{\rm ther})\rho^2(r_{\rm ther})=1,
\end{equation}
where $\kappa_{\rm abs}$ and $\kappa_{\rm sct}$ are the absorption and scattering opacities. The absorption opacity is determined as $k_{\rm abs}=k_{\rm R}-k_{\rm sct}$, with a Rosseland mean opacity, $k_{\rm R}$, given by the TOPS opacity tables \citep{Colgan16} and $k_{\rm sct}$ evaluated using the number of free electrons provided by the tables\footnote{Opacity and free electron number density tables were taken from http://aphysics2.lanl.gov/cgi-bin/opacrun/tops.pl.}. This choice of the mean absorption opacity gives a higher weight to frequency bands where the total cross-section is small, through which radiation more readily escapes. In contrast with RW11, who used pure H:He mixtures, we consider here plasma compositions with solar and 0.1 solar metallicity \citep{Asplund09}. We find that for $\Mc/\Menv\le1$, the ratio of $\Tcol$, obtained from the numerical calculations, to $\Tph$, given by eq.~(\ref{eq:RW11}), is $1.1[1.0]\pm0.05$ for $n=3/2[3]$ in the relevant temperature range. For large values of $\Mc/\Menv$, $\Mc/\Menv=10$, $\Tcol/\Tph$ is lower by $\approx10\%$. The fact that $\Tcol/T_{\rm ph}$ is close to unity suggests that the deviations from thermal spectra are not large, and that the spectral luminosity per unit wavelength $\lambda$ may be approximated by eq.~(\ref{eq:L_spec}), see \S~\ref{sec:analytic}.

The decrease of $\Teff$ below the self-similar result at $t\sim t_{\rm tr}$ is not reflected in the $\Tcol$ plot of fig.~\ref{fig:Tc32}, which shows the evolution only down to 0.7~eV. This is due to the fact that $T$ remains above 0.7~eV up to $t\sim t_{\rm tr}$ only for very low mass envelopes. For $n=3/2[3]$, $T_{\rm eff}(t=t_{\rm tr}/a)>0.7$~eV for
\begin{equation}\label{eq:Mevn_min_rec}
  \Menv<0.4 [3] \frac{R_{13}^{1.1} \vt_{*,8.5}}{\kappa_{0.34}^{2.1}}~M_\odot,
\end{equation}
well below the envelope masses used for the numerical calculations shown in fig.~\ref{fig:Tc32}. For low envelope masses and large stellar radii, the late time decrease of $T_{\rm eff}$, compared to the self-similar solution, is reflected in the color temperature at $T>0.7$~eV, as demonstrated in fig.~\ref{fig:Tc32_Menv1}.

\subsection{An analytic description of the post-breakout cooling emission}
\label{sec:analytic}

We provide here a summary of the analytic formulae which, based on the comparison of the numerical results with eqs.~\ref{eq:RW11}, provide an approximate description of the post-breakout cooling emission at times (see eq.~\ref{eq:t_limits})
\begin{equation}\label{t_lim1}
     t>0.2\frac{R_{13}}{\vt_{\rm s*,8.5}}\max\left[0.5,\frac{R_{13}^{0.4}}{(f_\rho\kappa_{0.34}M_0)^{0.2}\vt_{\rm s*,8.5}^{0.7}} \right]\,{\rm d}.
\end{equation}

The bolometric luminosity is described at early time by the self-similar expression (see eq.~\ref{eq:RW11})
\begin{equation}\label{eq:L1}
  \LRW =  2.0~[2.1]\times10^{42} \left(\frac{\vt_{\rm s*,8.5}t_{\rm d}^2}{f_\rho M_0\kappa_{0.34}}\right)^{-\epsilon_2} \frac{\vt^2_{\rm s*,8.5}R_{13}}{\kappa_{0.34}}\,{\rm \frac{erg}{s}},
\end{equation}
with $\epsilon_2=0.086[0.175]$ for $n=3/2[3]$. The luminosity is suppressed at late time by a factor, which may be approximated by
\begin{equation}\label{eq:Lq1}
  L/\LRW= A\exp\left[-\left(\frac{at}{t_{\rm tr}}\right)^\alpha\right],
\end{equation}
with $A=0.94[0.79]$, $a=1.67[4.57]$ and $\alpha=0.8[0.73]$ for $n=3/2[3]$. This approximation holds for a wide range of $\Rc/R$ and $\Mc/\Menv$ values, $10^{-3}\le\Rc/R\le10^{-1}$ and $0.1\le\Mc/\Menv\le10$, to better than $20[30]\%$ from $t\sim0.01t_{\rm tr}$ up to $t=t_{\rm tr}/a(n)$ for $n=3/2[3]$. $t=t_{\rm tr}$ is given by eq.~(\ref{eq:tstar}),
\begin{equation}\label{eq:tstar1}
  t_{\rm tr}=\left(\frac{\kappa \Menv}{8\pi c\vt_{\rm s*}}\right)^{1/2}=19.5\left(\frac{\kappa_{0.34} M_{\rm env,0}}{\vt_{\rm s*,8.5}}\right)^{1/2}{\rm d}.
\end{equation}

The spectral luminosity per unit wavelength $\lambda$ may be approximated by (RW11)
\begin{equation}\label{eq:L_spec}
  L_\lambda(t)\equiv\frac{dL}{d\lambda}=L(t)\frac{\Tcol}{hc}g_{\rm BB}(hc/\lambda\Tcol),
\end{equation}
where $g_{\rm BB}$ is the normalized Planck function,
\begin{equation}\label{eq:gBB}
  g_{\rm BB}(x)= \frac{15}{\pi^4}\frac{x^5}{e^x-1},
\end{equation}
and $\Tcol$ is given by $\Tcol/\Tph=1.1[1.0]\pm0.05$ for $n=3/2[3]$ with weak sensitivity to metallicity in the relevant temperature range (for large radii, $R>10^{13.5}$~cm, and large values of $\Mc/\Menv$, $\Mc/\Menv=10$, $\Tcol/T_{\rm ph}$ is lower by $\approx10\%$; see figs.~\ref{fig:Tc32} and~\ref{fig:Tc3}). $\Tph$ is given by eq.~(\ref{eq:RW11}),
\begin{equation}\label{eq:T1}
  \Tph = 1.61[1.69]\left(\frac{\vt^2_{\rm s*,8.5}t_{\rm d}^2}{f_\rho M_0\kappa_{0.34}}\right)^{\epsilon_1} \frac{R_{13}^{1/4}}{\kappa^{1/4}_{0.34}}t_{\rm d}^{-1/2}\,{\rm eV},
\end{equation}
with $\epsilon_1=0.027[0.016]$ for $n=3/2[3]$.

The dependence of the results on $f_\rho$ is weak. For $\Rc/R\ll1$, $f_\rho$ may be approximated by $f_\rho=(\Menv/\Mc)^{1/2}$ and $f_\rho=0.08(\Menv/\Mc)$ for $n=3/2$ and $n=3$ respectively (for progenitors with $n=3$ envelopes and large core radii, $\Rc/R\approx0.1$, $f_\rho$ is larger by a factor of $\approx3$ than the value given by this approximation, see fig.~\ref{fig:frho}).

The above results are valid for $T>0.7$~eV, i.e. for
\begin{equation}\label{eq:t_T}
  t<7.4\left(\frac{R_{13}}{\kappa_{0.34}}\right)^{0.55} \, {\rm d}.
\end{equation}

\section{Double-peaked SN light curves}
\label{sec:2peak}

\begin{figure}[h]
\epsscale{1} \plotone{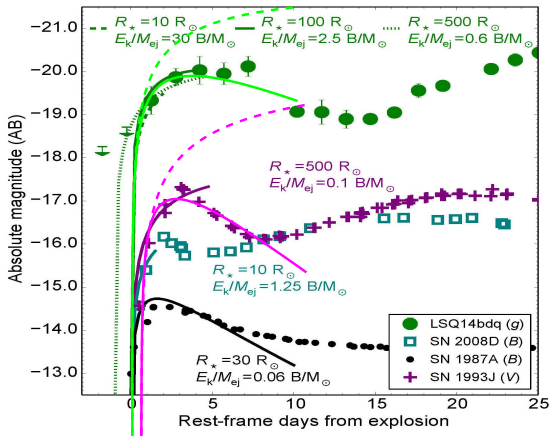}
\caption{Double peaked SN light curves. Solid (dashed) light green/magenta lines are derived from eqs.~(\ref{eq:RW11}) with (without) the suppression of eq.~(\ref{eq:Lq}), and are overlayed on a figure adapted from \citet{Nicholl15bdq} (the new curves extend outside of the y-axis range of the original figure). Solid green[magenta] lines reproduce the observed first peak of LSQ14bdq[1993J] for $n=3/2$, $R=10^{13.5}\kappa_{0.34}$~cm, $\vt_{\rm s*}=2.4[1.1]\times10^{9}{\rm cm/s}$ ($E/M\approx f_\rho^{2\beta_1}\vt_{\rm s*}^2=8[2]\times10^{51}{\rm erg}/M_\odot$), and $\Menv=1.3[0.11]\kappa_{0.34}^{-1}M_\odot$. The light curves extend up to where T=0.7eV, as determined by eq.~(\ref{eq:t_T}) (the 1993J explosion time is shifted by 0.5~d compared to the choice of the original figure; We do not show models for SNe 1987A and 2008D).
\label{fig:2peak}}
\end{figure}

The bolometric light curves of several SNe, mainly of the IIb class \citep{Wheeler93J,Arcavi11dh,VanDyk14df} but also some super-luminous SN of type I \citep[see][for a recent discussion]{Nicholl15dpSL}, show a "double peak" behavior: a first peak at a few days after the explosion, preceding the main SN peak (on time scale of tens of days). It is commonly accepted that the first peak is produced by the post-breakout shock cooling radiation from an extended, $R\sim10^{13}$~cm, low mass, $M\le0.1M_\odot$ envelope \citep{Woosley9493J,Bersten12dh,NakarPiro14,Piro15}, which becomes transparent after a few days of expansion, and it is often argued that this extended envelope should be characterized by a non-standard structure, e.g. where the the mass is initially concentrated at $r\sim R$ \citep[e.g.][]{NakarPiro14,Piro15} \citep[alternatively, proponents of the central engine magnetar models for super-luminous SNe suggested that the first peak in such double-peaked SNe may due to shock breakout from ejecta that was inflated to large radius by the energy output of the magnetar, e.g.][]{Kasen15SLdp}.

We find that the suppression of $L$ at $t=t_{\rm tr}/a$ may naturally account for double-peaked SN light curves, with a first peak obtained on a few days time scale for $\Menv<1M_\odot$, without a need for non-standard structure. This is demonstrated in fig.~\ref{fig:2peak}, where the first peaks of some prototypical double-peaked light curve SNe, LSQ14bdq and 1993J, are reproduced by the post-breakout emission described by eqs.~(\ref{eq:RW11}) and~(\ref{eq:Lq}) with $n=3/2$, $T_{\rm col}/\Tph=1.1$, $f_\rho=0.3$ (as may be appropriate for large $\Mc/\Menv)$, $R=10^{13.5}~{\rm cm}\kappa_{0.34}, \vt_{\rm s*}=2.4[1.1]\times10^{9}{\rm cm/s}$ ($E/M\approx f_\rho^{2\beta_1}\vt_{\rm s*}^2=8[2]\times10^{51}{\rm erg}/M_\odot$), and $\Menv=1.3[0.11]M_\odot$ for LSQ14bdq[1993J]. We note, that the suppression of the bolometric luminosity at $t\ge t_{\rm tr}/a$ is determined mainly by $\Menv/\vt_*$ and is not sensitive to the values of $\Rc/R$ and of $\Mc/M$.

The late-time spectrum of LSQ14bdq (taken well after the first peak) shows no evidence for Hydrogen, indicating that our analysis, which is valid for Hydrogen-rich envelopes, may not directly apply to this case (a detailed derivation of an upper limit on the Hydrogen mass in LSQ14bdq is yet to be performed). However, our analysis does provide an approximate description of the emission from a Helium dominated envelope at the relevant times. At $1~{\rm eV} \lesssim T \lesssim 1.7~{\rm eV}$ (and the relevant density range) Helium remains singly-ionized and its opacity may be approximated by $\kappa=0.1~{\rm cm}^2 {\rm g}^{-1}$. Thus, using $\kappa=0.1~{\rm cm}^2 {\rm g}^{-1}$ our results provide an approximate description of the emission from a He envelope up to $\sim5$~d. At later time the optical flux will continue to decrease due to the decrease in bolometric luminosity. The time dependence of the flux is not expected to be accurately described by our model at late time, due to the variation of the opacity.

We did not carry out a detailed analysis of the allowed range of model parameters, as our main goal was to demonstrate that the suppression of the bolometric luminosity is consistent with a polytropic envelope, and since the two observable quantities (peak time and luminosity) constrain, but do not enable an accurate determination of, the model parameters, $\{\Menv,\vt_{\rm s*},R/\kappa\}$, which determine these observable quantities. In particular, the relation between $t_{\rm tr}$, which is determined mainly by $\Menv/\vt_{\rm s*}$ (see eq.~\ref{eq:tstar}), and the peak time, $\approx t_{\rm tr}/a$, depends on the envelope structure through the dependence of $a$ on $n$. The variation of $a$ from $\approx 2$ to $\approx 4$ between $n=3/2$ and $3$ implies a factor $\sim4$ uncertainty in inferring $\Menv/\vt_{\rm s*}$ (in the absence of additional constraints on the envelope structure).

The values of $R$, $M_{\rm env}$ and $\vt_{\rm s*}$ inferred by our analysis are consistent with those inferred by \cite{Woosley9493J} for SN 1993J and by \cite{Piro15} for SN 1993J and LSQ14bdq. Our results show that a non-standard envelope structure \citep[with extended envelope mass concentrated around the outer radius $R$, e.g. \S~3 and fig. 2 of][]{NakarPiro14} is not required.

\section{Discussion}
\label{sec:discussion}

\subsection{Early time, $t<t_\delta=$few days}

We have used numerical calculations to demonstrate that the early, $t<t_\delta=$few days (see eq.~(\ref{eq:t_limits})), envelope cooling emission is not sensitive to the details of the density profile of the envelope (see figs.~\ref{fig:Teff32}-\ref{fig:Lq3}). The emission is well described by eqs.~(\ref{eq:RW11}), with $T_{\rm ph}$ determined mainly by $R$, and $L$ determined mainly by $\vt_{\rm s*}^2R$. For $\Mc/\Menv\le1$, the ratio of $\Tcol$ (see \S~\ref{sec:rad}, eqs.~(\ref{eq:L_spec},\ref{eq:gBB})), obtained from the numerical calculations, to $T_{\rm ph}$, given by eq.~(\ref{eq:RW11}), is $1.1[1.0]\pm0.05$ for $n=3/2[3]$, with weak sensitivity to metallicity in the relevant temperature range (this value is somewhat lower than that obtained in RW11, 1.2, who considered a pure He:H mixture; for large radii, $R>10^{13.5}$~cm, and large values of $\Mc/\Menv$, $\Mc/\Menv=10$, $\Tcol/T_{\rm ph}$ is lower by $\approx10\%$; see figs.~\ref{fig:Tc32} and ~\ref{fig:Tc3}).

The weak dependence of the early emission on the density structure, reflected in the very weak dependence of $\Tcol/T_{\rm ph}$ and of $L$ and $T$ in eqs.~(\ref{eq:RW11}) on $n$ and model parameters other than $R$ and $\vt_{\rm s*}^2$, implies that $R$ and $\vt_{\rm s*}^2$ may be inferred accurately and robustly from the observations of the early UV/optical emission.

The approximate relation between $\vt_{\rm s*}$ and $E/M$, given by eq.~(\ref{eq:vstar}), holds to better than 10\% for $0.3<\Mc/M_{\rm env}<3$ (see fig.~\ref{fig:vst}). The dependence of $f_\rho$ on $n$ and on $\Mc/M_{\rm env}$, approximately given $\Rc/R\ll1$ by $f_\rho=(\Menv/\Mc)^{1/2}$ and $f_\rho=0.08(\Menv/\Mc)$ for $n=3/2$ and $n=3$ (see fig.~\ref{fig:frho}), implies that the relation between $\vt_{\rm s*}$ and $E/M$ depends on the ejecta structure. $E/M$ may be inferred from $\vt_{\rm s*}$ by $E/M=0.9[0.3]\vt_{\rm s*}^2$ for $n=3/2[3]$ with $5[30]\%$ accuracy for $0.3<\Mc/M_{\rm env}<3$ (for progenitors with $n=3$ envelopes and large core radii, $\Rc/R\approx0.1$, $f_\rho$ is larger and $E/M=0.5\vt_{\rm s*}^2$ is a better approximation; see fig.~\ref{fig:frho}). Conversely, a comparison of $\vt_{\rm s*}$, determined by early UV observations, and $E/M$, determined by other late time observations (e.g. spectroscopic ejecta velocity), will constrain the progenitor structure.

\subsection{Late time, $t>t_\delta$}

We have extended the solutions to $t\sim t_{\rm tr}$, see eq.~(\ref{eq:tstar}), when the emission emerges from deep within the envelope and depends on the progenitor's density profile. The expression given in the abstract for $t_{\rm tr}$ is obtained from eq.~(\ref{eq:tstar}) using the approximation of eq.~(\ref{eq:vstar}) for $\vt_{\rm s*}$, dropping for the sake of simplicity the dependence on $f_\rho$, $t_{\rm tr} = 13f_\rho^{\beta/2}(\Menv/M_\odot)^{3/4}(M/\Menv)^{1/4}(E/10^{51}{\rm erg})^{-1/4}$~d.

We have shown (see \S~\ref{sec:LateForm}) that the dependence of $L$ and $T$ on the progenitor parameters is of the general form of eq.~(\ref{eq:T_L_dimensions}), and used the numerical solutions to determine the dimensionless functions $f_T$ and $f_L$ for polytropic, $n=3/2$ and $n=3$, envelopes with a wide range of core to envelope mass and radius ratios, $0.1<\Mc/(M-\Mc)<10$, $0.001<\Rc/R<0.1$. We have found that $T$ is well described by the analytic solution also at late time (for low mass envelopes $T$ may drop at late time by $\sim20\%$ below the analytic prediction, see eq.~\ref{eq:Mevn_min_rec} and fig.~\ref{fig:Tc32_Menv1}), while $L$ is suppressed by a factor which depends mainly on $n$ (and only weakly on $\Rc/R$ and $\Mc/M$), and may be approximated to $\approx20\%$ accuracy up to $t=t_{\rm tr}/a(n)$ by the analytic approximations of eq.~(\ref{eq:Lq}).

For very large progenitors, $R>10^{13.5}$~cm, with low mass envelopes, $\Menv\le1M_\odot$, the separation of the time scales $R/\vt_{\rm s*}$ and $t_{\rm tr}/a$ is not large, and the analytic expression for $L$ given by eqs.~(\ref{eq:RW11}), which holds for $R/\vt_{\rm s*}\ll t\ll t_{\rm tr}/a$, is not accurate at any time. However, as demonstrated in fig.~\ref{fig:Lq32}, the approximation for $L$ obtained using eqs.~(\ref{eq:RW11}) with the suppression factor of eq.~(\ref{eq:Lq}) is accurate to better than $10\%$ up to $t=0.1t_{\rm tr}$ also in this case. This implies that $R\vt_{\rm s*}^2$ (and hence $\vt_{\rm s*}^2$) may be accurately determined from the bolometric luminosity $L$ at early time also for very large progenitors, $R>10^{13.5}$~cm, with low mass envelopes.

It is worth noting, that the suppression of $L$ at $t>t_\delta$ implies that using eqs.~(\ref{eq:RW11}) to infer $R$ from the luminosity observed at $t>t_\delta$ would lead to an under estimate of $R$ due to the overestimate of $L$, as demonstrated in fig.~\ref{fig:2peak} (compare the solid and dashed curves) and as discussed by \citet{Rubin15}.

We have shown (see fig.~\ref{fig:2peak}) that the suppression of $L$ at $t_{\rm tr}/a(n)$ obtained for standard polytropic envelopes may account for the first optical peak of double-peaked SN light curves, with first peak at a few days for $\Menv<1M_\odot$. The suppression of the bolometric luminosity is consistent with the observed behavior, and does not require a non-polytropic envelope with a special structure, e.g. where the the mass is initially concentrated at $r\sim R$. The time at which the bolometric luminosity is suppressed corresponds to $t_{\rm tr}/a(n)$ and hence constrains $\Menv/\vt_{\rm s*}$ (see eq.~(\ref{eq:tstar})), while the luminosity constrains $\vt_{\rm s*}^2R$. It is important to emphasize that these parameters cannot be determined accurately from the observations, since the emission at $t>t_\delta$ depends on the detailed structure of the progenitor (see discussion at the end of \S~\ref{sec:2peak}).

Finally, it is important to emphasize that our analysis holds as long as the opacity is approximately that of a fully ionized gas, i.e. for $T>0.7$~eV, $t<14R_{13.5}^{0.55}$~d. At lower temperatures, recombination leads to a strong decrease of the opacity (see fig.~\ref{fig:opacity}) and the photosphere penetrates deep into the ejecta, to a depth where the temperature is sufficiently high to maintain significant ionization and large opacity, implying that $T$ does not drop significantly below $\sim 0.7$~eV. This enhances the dependence on the details of the envelope structure and implies that detailed radiation transfer models are required to describe the emission (our simple approximations for the opacity no longer hold).

\subsection{The importance of early UV observations}

An accurate determination of $R$ requires an accurate determination of $T$ at a time when eq.~(\ref{eq:RW11}) holds and $T$ depends mainly on $R$, i.e. when $T>0.7$~eV. An accurate determination of $T$ requires, in turn, observations at $\lambda<hc/4T=0.3(T/1{\rm eV})^{-1}\mu$, in order to identify the peak in the light curve, which is obtained when $T$ crosses $T_\lambda\approx hc/4\lambda$ (or by identifying the spectral peak provided redenning can be corrected for, RW11). Since the emission peaks below $0.3\mu$ for $T>1$~eV, UV observations at $\lambda<0.3~\mu$ (which must be carried out from space) will enable one to reliably determine $T$ and $R$ (and hence also $\vt_{\rm s*}$). On the other hand, observations at $\lambda\ge0.44\mu$ (B-band or longer) corresponding to $T_\lambda=hc/4\lambda\le0.7$~eV, will not enable one to accurately determine $T$ and $R$. Observations in the U-band, $\lambda=0.36\mu$ corresponding to $hc/4\lambda=0.8$~eV, will provide less accurate results than UV observations due to the strong temperature dependence of the opacity at slightly lower temperature.

\acknowledgements We thank A. Rubin, B. Katz, E. Ofek and A. Gal-Yam for useful discussions and constructive comments. This research was partially supported by an ISF I-Core, an IMOS grant and a Minerva grant.

\bibliographystyle{hapj}
\bibliography{SNbreakout}
\end{document}